%Paper: hep-th/9506198
%From: "Andre LeClair" <leclair@hepth.cornell.edu>
%Date: Thu, 29 Jun 1995 16:09:03 -0400

%

\input harvmac.tex      %% input your version
%\input epsf.tex         %%        "
%\input harvmac
%%%%%%%%%%%%%%%%%%%%%%%%%%%%%%%%%%%%%%%%%%%%%%%%%%%%%%%%%%%%%%%%%%%%%
%%%%
%%%%
%%%%       Affine Lie Algebra Symmetry of Sine-Gordon Theory
%%%%		  at Reflectionless Points.
%%%%
%%%%               By Andre LeClair and  Dennis Nemeschansky
%%%%
%%%%
%%%%%%%%%%%%%%%%%%%%%%%%%%%%%%%%%%%%%%%%%%%%%%%%%%%%%%%%%%%%%%%%%%%%%
%%%%%%%
%%%%            Macros needed:  harvmac.tex (from bulletin board)
%%%%

%%%%%%%%%%%%%%%%%%%%%%%%%%%%%%%%%%%%%%%%%%%%%%%%%%%%%%%%%%%%%%%
%
%		DEFINITIONS FOR TEX
%
%%%%%%%%%%%%%%%%%%%%%%%%%%%%%%%%%%%%%%%%%%%%%%%%%%%%%%%%%%%%%%%
%

\def\tilde{\widetilde}
\def\bar{\overline}
\def\hat{\widehat}
\def\*{\star}
\def\[{\left[}
\def\]{\right]}
\def\({\left(}		
\def\){\right)}

%
%%%%%%%%%%%%%%%%%%%%%%%%%%%%%%%%%%%%%%%%%%%%%%%%%%%%%%%%%%%%%%%
%
\def\zb{{\bar{z} }}
\def\frac#1#2{{#1 \over #2}}
\def\inv#1{{1 \over #1}}

\def\d{\partial}

\def\2pi{\hbox{$2\pi i$}}

\def\dsl{\raise.15ex\hbox{/}\kern-.57em\partial}
\def\Dsl{\,\raise.15ex\hbox{/}\mkern-.13.5mu D}
%
%%%%%%%%%%%%%%%%%%%%GREEK LETTERS%%%%%%%%%%%%%%%%%%%%%%%%%%%%%%
%

%
%%%%%%%%%%%%%%%%%%%CALIGRAPHIC LETTERS%%%%%%%%%%%%%%%%%%%%%%%%%
%

		\def\CO{{\cal O}}

	\def\CZ{{\cal Z}}

\def\2pi{\hbox{$2\pi i$}}

\def\dsl{\raise.15ex\hbox{/}\kern-.57em\partial}
\def\Dsl{\,\raise.15ex\hbox{/}\mkern-.13.5mu D}
%
%%%%%%%%%%%%%%%%%%%%GREEK LETTERS%%%%%%%%%%%%%%%%%%%%%%%%%%%%%%
%
%%%%%%%%%%%%%%% MATH CHARACTERS %%%%%%%%%%%%%%%%%%%%%%%%%%%%
%
\font\numbers=cmss12
%\font\numbers=cmu10 scaled\magstep1
\font\upright=cmu10 scaled\magstep1
\def\stroke{\vrule height8pt width0.4pt depth-0.1pt}
\def\topfleck{\vrule height8pt width0.5pt depth-5.9pt}
\def\botfleck{\vrule height2pt width0.5pt depth0.1pt}
\def\Zmath{\vcenter{\hbox{\numbers\rlap{\rlap{Z}\kern
0.8pt\topfleck}\kern
2.2pt
                   \rlap Z\kern 6pt\botfleck\kern 1pt}}}
\def\Qmath{\vcenter{\hbox{\upright\rlap{\rlap{Q}\kern
                   3.8pt\stroke}\phantom{Q}}}}
\def\Nmath{\vcenter{\hbox{\upright\rlap{I}\kern 1.7pt N}}}
\def\Cmath{\vcenter{\hbox{\upright\rlap{\rlap{C}\kern
                   3.8pt\stroke}\phantom{C}}}}
\def\Rmath{\vcenter{\hbox{\upright\rlap{I}\kern 1.7pt R}}}
\def\Z{\ifmmode\Zmath\else$\Zmath$\fi}
\def\Q{\ifmmode\Qmath\else$\Qmath$\fi}
\def\N{\ifmmode\Nmath\else$\Nmath$\fi}
\def\C{\ifmmode\Cmath\else$\Cmath$\fi}
\def\R{\ifmmode\Rmath\else$\Rmath$\fi}
%%%%%%%%%%%%%%%%%%%%%%%%%%%%%%%%%%%%%%%%%%%%%%%%%%%%%%%%%%%%%%%%%
 %%%%%%%%%%%%%%%%%% END OF DEFINITIONS %%%%%%%%%%%%%%%%%%%%%%
 %%%%%%%%%%%%%%%%%%%%%%%%%%%%%%%%%%%%%%%%%%%%%%%%%

\Title{CLNS 95/1340 \ \ CERN-TH/95-113}
{\vbox{\centerline{Affine Lie Algebra Symmetry of Sine-Gordon Theory  }
\centerline{ at Reflectionless Points} }}

\bigskip
\bigskip

\centerline{Andr\'e LeClair}
\centerline{Newman Laboratory}
\centerline{Cornell University}
\centerline{Ithaca, NY  14853}
\bigskip\bigskip

\centerline{Dennis Nemeschansky$^{*}$}
\footnote{}{$^{*}$ On leave from  Physics Department,
University of Southern California, Los Angeles, CA 90089}
\centerline{Theory Division, CERN }
\centerline{CH-1211, Geneva 23, Switzerland}

\vskip .3in
The quantum affine symmetry of the sine-Gordon theory at $q^2=1$,
which occurs at the reflectionless points, is studied.
Conserved currents that correspond to the closure of simple root generators
are considered, and shown to be local.
We argue that they satisfy the $\hat {sl(2)}$ algebra.
Examples of these currents are explicitly constructed.

\Date{5/95}
%\draftmode
%
%
%
%
%
%
%      pecdef
\noblackbox

\def\zb{{\bar{z}}}

%
%
%
%
%sample reference
%
%on an operator formulation of the superstring\ref\a{\sdual} .
%
%sample equations
%\eqn\one{
%\V 123 A\rangle_1 \vb_2 \vc_3 =
%\langle h_1 \left[ V_A (0)\right] h_2 \left[ V_B (0) \right]
%h_3 \left[ V_C (0) \right] \rangle . }
%
%
% \eqn\two{\eqalign{  a & = b \cr c & = d \cr}}
%
%\eqnn\two
%$$\eqalignno{
% h_1 (z ) &= z \cr
 %h_2 (z ) &= { 1\over {1-z } }&\two\cr
% h_3 (z ) &= { {z -1}\over {z} } .\cr}$$
%
%eq a,b,c etc
%\eqna\three
%$$\eqalignno{
%.............&\three {a} \cr
%..........&\three {b} \cr
%}$$
%\eqn\number{\eqalign{  ......... \cr}}       numbers automatically

\newsec{Introduction }

It is known that many massive integrable quantum field theories
possess dynamical symmetries corresponding to $q$-deformations
of affine Lie algebras, and that these symmetries place
powerful constraints on the S-matrices.  In \ref\rbl{D. Bernard
and A. LeClair, Commun. Math. Phys. 142 (1991) 99.}, the
conserved currents which correspond to the finite number of
simple roots of the $q$-deformed affine Lie algebra were
constructed using conformal perturbation theory, as developed
by Zamolodchikov\ref\rzamo{A. B. Zamolodchikov, Int. J. Mod.
Phys. A4 (1989) 4235.}.  The $q$-deformed
affine Lie algebra has an infinite number of generators,
and a presentation of the complete set of algebraic relations
was obtained by Drinfel'd\ref\rdrin{V. G. Drinfel'd,
Sov. Math. Dokl. Vol. 36 (1988) 212.}.   It is therefore
interesting to investigate the properties of the conserved currents
for the infinite number of additional generators which arise
upon closure of the simple-root generators.  For generic $q$
this is expected to be very complicated, since the currents for
the simple roots have fractional Lorentz spin, and $q^2$ is
a braiding phase.  This implies that in general the braided
commutators of the simple root generators  do not close on
charges which are integrals of local currents.  Thus, for
generic $q$, the currents for the higher Drinfel'd charges
are expected to be highly non-local.

In this paper we study the sine-Gordon (SG) theory when $q^2=1$,
which occurs at the so-called reflectionless points of the
SG coupling $\beta$.  At these points, the locality properties
of the currents for the simple roots are such that the
conserved currents for the non-simple roots are local.  Our
main results are the following.  We conjecture the existence
of an affine $\hat{sl(2)}$ symmetry at all of the reflectionless
points of the SG theory.  This does not follow directly from
the $q^2=1$ specialization of the results in \rbl, since, as we
will see, the complete relations of the affine Lie algebra necessarily
arise at higher order in conformal perturbation theory.  The
difficulty in carrying out explicitly higher order conformal
perturbation theory is what limits us to only making a conjecture
about the existence of the $\hat{sl(2)}$ symmetry.
We also construct explicitly some of the higher currents and show that
they are products of the currents for the simple roots with local
operators.

\newsec{$\hat{sl(2)}$ Symmetry at Reflectionless Points}

\def\bh{\hat{\beta}}
\def\ph{\varphi}
\def\phb{\bar{\varphi}}
\def\sl{\hat{sl(2)}}

We consider the SG theory defined by the Euclidean action
\eqn\eIIi{
S = \inv{4\pi}  \int d^2 z \( \d_z \Phi \d_\zb \Phi + 4 \lambda
\cos  (\bh \Phi ) \) , }
where $z, \zb$ are the usual Euclidean light-cone coordinates,
and $\bh$ is related to the conventional SG coupling $\beta$ as
$\bh = \beta/{\sqrt{4\pi}}$.
The conserved $U(1)$ topological charge normalized to be $\pm 1$
on the solitons is
\eqn\eIIii{
T_0 = \frac{\bh}{2\pi} \int_{-\infty}^{\infty}  dx \d_x \Phi . }

In \rbl, four additional charges corresponding to the simple root
generators of the $q-\hat{sl(2)}$ were constructed in conformal
perturbation theory:
\eqn\eIIiii{\eqalign{
Q^\pm_{-1}  &= \int dz \> J^\pm_{(-1)} + \int d\zb \> \bar{J}^\pm_{(-1)}
\cr
Q^\pm_{1}   &= \int dz \> J^\pm_{(1)} + \int d\zb \> \bar{J}^\pm_{(1)}
\cr
}}
where
\eqn\eIIiv{
\eqalign{
J^\pm_{(-1)} &= \exp \(  \pm \frac{2i}{\bh} \ph \)
{}~~~~~~~
\bar{J}^\pm_{(-1)} = \lambda \frac{\bh^2}{\bh^2 -2}
\exp \[ \pm i (\frac{2}{\bh} - \bh )\ph \mp i\bh \phb \]
\cr
\bar{J}^\pm_{(1)} &= \exp \(  \mp \frac{2i}{\bh} \phb \)
{}~~~~~~~
J^\pm_{(1)} = \lambda \frac{\bh^2}{\bh^2 -2}
\exp \[ \mp i (\frac{2}{\bh} - \bh )\phb \pm i\bh \ph \]
\cr}}
and $\ph , ~\phb$ are quasi-chiral components of $\Phi$ defined in
conformal perturbation theory:  when $\lambda = 0$,
$\Phi = \ph (z) + \phb (\zb )$.   For generic $\bh$, these currents
are exact to first order in $\lambda$.
Together with the topological charge $T_0$,
$Q^\pm_{1} , ~ Q^\pm_{-1} $ satisfy the defining relations for the
simple roots generators of the $q-\sl$ algebra, where
$q= \exp(-2\pi i/\bh^2 )$.  In particular
one has
\eqn\eIIvi{
Q^\pm_{-1} Q^\mp_1  - q^{-2} Q^\mp_1 Q^\pm_{-1}
= \frac{\lambda}{2\pi i}
\( \frac{\bh^2}{2-\bh^2} \)^2
\( 1 - q^{\pm 2 T_0 } \) . }

\def\qt{\tilde{Q}}
\def\tt{\tilde{T}}

The $q$ deformation parameter arises as a braiding phase for the above
currents.  This is related to the fact that the charges have fractional
Lorentz spin in general.  Let $L$ denote the generator of Euclidean
rotations.  Then,
\eqn\eIIv{
\[ L, Q^\pm_{-1} \] = \( \frac{2}{\bh^2} - 1 \) ~ Q^\pm_{-1}
,~~~~~~~
\[ L, Q^\pm_{1} \] = - \( \frac{2}{\bh^2} - 1 \) ~ Q^\pm_{1}
}

We now specialize to the reflectionless points:
\eqn\eIIvii{
\bh^2 = \frac{2}{N+1} }
where $N$ is a positive integer.  The point $N=1$ corresponds
to the free fermion point.  For each of these couplings
$q^2 = 1$ and the Lorentz spin of the above charges is $\pm N$.

Since $q^2 = 1$, one is led to consider whether an undeformed
affine $\sl$ symmetry exists in these models.   The affine
$\sl$ algebra in the principal gradation is generated by
$\qt^\pm_n$, for $n$ and odd integer, and $\tt_n$ for $n$ even,
satisfying
\eqn\eIIviii{
\[\tt_n , \tt_m \] = 0,~~~~~
\[\tt_n , \qt^\pm_m \] = \pm 2 ~ \qt^\pm_{n+m} ,
{}~~~~~~
\[ \qt^+_n , \qt^-_m \] = \tt_{n+m}  . }

To understand how this structure can arise at the reflectionless
points, consider first the conformal limit $\lambda = 0$, where
the charges $Q^\pm_{-1}$ become left-moving and $Q^\pm_1$ become
right-moving.  When $q^2=1$, $Q^\pm_{-1} $ satisfy the undeformed
Serre relations for $\sl$, and similarly for $Q^\pm_1$\foot{The Serre
relations were proven for arbitrary $q$ in \ref\rfl{G. Felder and
A. LeClair, Int. J. Mod. Phys. A7 Suppl. 1A (1992) 239. }}.
The commutator algebra of $Q^\pm_{-1}$ closes on a set of left-moving
charges $Q^\pm_{-n} , ~ T_{-n}, ~ n>0$, and similarly
$Q^\pm_1$ closes on right-moving charges $Q^\pm_n , ~T_n , ~n>0$.
Due to the well-behaved locality properties at the reflectionless
points, one can in principal explicitly compute the currents for
these higher charges.   When $\lambda =0$, since the left and right
moving charges commute, one thus obtains two decoupled Borel
subalgebras of $\sl$.

Consider now turning on the perturbation $\lambda \neq 0$.  Since
$Q^\pm_1 , ~ Q^\pm_{-1}$ are conserved to first order in perturbation
theory, then so are the higher charges.  Therefore one obtains
\eqn\eIIix{\eqalign{
Q^\pm_{n}  &= \int dz \> J^\pm_{(n)} + \int d\zb \> \bar{J}^\pm_{(n)}
\cr
T_n    &= \int dz \> J_{(n)} + \int d\zb \> \bar{J}_{(n)}
\cr
}}
where when $\lambda = 0$, $\bar{J}^\pm_{n<0} = \bar{J}_{n<0} = 0$,
and $J^\pm_{n>0} = J_{n>0} = 0$.
When $\lambda \neq 0$, the two Borel subalgebras no longer commute,
and the interesting question is whether they do indeed satisfy the
$\sl$ relations.  Consider for example the relation \eIIvi.  For
generic values of $\bh$, it was shown in \rbl\ that the RHS of
\eIIvi\ is exact to first order in $\lambda$.  When $q^2=1$, the
RHS of \eIIvi\ actually vanishes on states of integer topological
charge.

A simple scaling argument shows that the complete
$\sl$ relations must arise at higher order in perturbation theory.
 From the Lorentz spin of $Q^\pm_1 , ~ Q^\pm_{-1}$, one deduces that
\eqn\eIIx{
\[ L , Q^\pm_n \] = -n N ~ Q^\pm_n  , ~~~~~
\[L, T_n \] = -nN ~ T_n .  }
In the conformal limit, $L= L_0 - \bar{L}_0 $ where
$L_0$ and $\bar{L}_0$ are the Virasoro zero modes, and the scaling
dimensions of operators are given by $L_0 + \bar{L}_0$.  Since
in the conformal limit the negative (positive) frequency modes
of $\sl$ are left (right) moving, one deduces  the following
scaling dimension of the charges:
\eqn\eIIxi{
{\rm dim} \( Q^\pm_n \) = {\rm dim} \( T_n \) = N |n|.  }
The parameter $\lambda$ has scaling dimension $2-\bh^2 = 2N/(N+1)$,
thus
$\qt^\pm_n = (c\lambda)^{-|n|(N+1)/2} Q^\pm_n $  and
$\tt_n = (c\lambda)^{-|n|(N+1)/2} T_n $  are dimensionless operators,
where $c$ is a dimensionless constant.  Taking the latter operators
to satisfy the algebra \eIIviii, one obtains
\eqn\eIIxii{
\eqalign{
\[ T_n , T_m \] &= 0  \cr
\[ T_n , Q^\pm_m \] &= \pm 2
{}~
\( c\lambda \)^{(N+1)(|n| + |m| - |n+m| )/2 }
{}~
Q^\pm_{n+m}  \cr
\[ Q^+_n , Q^-_m \] &=
\( c\lambda \)^{(N+1)(|n| + |m| - |n+m| )/2 }
{}~ T_{n+m} . \cr }}
The powers of $\lambda$ on the RHS of \eIIxii\ are always integers,
and as $\lambda$ goes to zero, one obtains two decoupled Borel
subalgebras.

At the free fermion point $N=1$, the conserved charges satisfying
\eIIxii\ were constructed explicitly\ref\ral{A. LeClair, Nucl. Phys.
B415 (1994) 734.}.
In the next section we will provide evidence for the structure
\eIIxii\ by verifying the existence of some higher dimensional
currents to first order in $\lambda$.

Though we are unable to verify explicitly the algebra \eIIxii\ since
it involves higher order conformal perturbation theory, we can
at least verify that this structure is allowed in perturbation theory.
Consider the commutator $[Q^+_{-1} , Q^-_1 ]$.  In general one has
the following perturbative expansion:
\eqn\eIIxiii{
\[ Q^+_{-1} , Q^-_1 \]  = \sum_n \lambda^n
\( \int dz   \CO_n  + \int d\zb \bar{\CO}_n  \). }
In conformal perturbation theory $\CO_n$  must be a product
of $\exp (2i\Phi/\bh )$ with some power of the perturbation
$\cos (\bh \Phi )$.  Thus, $\CO_n $ must be the operator
$\exp ( (i(2/\bh + k\bh ) \Phi )$ or its derivatives for
$k\in \CZ$.  If $\CO_n$ involves $m$ derivatives, then the
dimension of $\CO_n$ is $(2/\bh + k\bh )^2  + m$.  Since the
dimension of $Q^+_{-1} , ~ Q^-_1 $ is $2/\bh^2 -1$ and that of
$\lambda$ is $2-\bh^2$, simple dimensional analysis requires
\eqn\eIIxiv{
2\( \frac{2}{\bh^2} - 1 \)  = n (2-\bh^2 ) + \( \frac{2}{\bh} + k\bh \)^2
+ m - 1  .}
For generic irrational $\bh$, the above equation has no solution
except for $n=1$.  However, at the reflectionless points, one
has the additional solution $n=N+1 , k= -(N+1), m=1$.  This corresponds
to the relation
\eqn\eIIxv{
\[ Q^+_{-1} , Q^-_1 \] =  (c\lambda)^{N+1} ~ T_0 , }
where
$T_0$ is the $U(1)$ charge.

\newsec{Higher Conserved Currents}

In this section we construct explicitly some of the higher conserved
currents at the reflectionless points.  One way of verifying the
structure proposed in the last section is as follows.  We focus
on the charges $Q^+_{-n}$ for $n>0$.  When $\lambda = 0$, the
current has one chiral component $J^+_{(-n)}$.  Given that $J^+_{-n}$
must have $U(1)$ charge $+2$, one considers the following form for
this current:
\eqn\eIIIi{
J^+_{(-n)}  = \exp \( \frac{2i}{\bh} \ph \) ~ \CO_{-n} (z) }
where
$\CO_{-n} $ is a local operator depending on powers of $\ph$ and its
derivatives.   At the reflectionless points, since the spin of
$Q^+_{-n}$ is $nN$, and that of $\exp (2i\ph /\bh )$ is $N+1$,
the spin of $\CO_{-n}$ must be $N(n-1)$.

The specific computation we carried out is the following.
Suppose one takes a current of the form  $J= \exp( 2i\ph /\bh )
\CO$ where $\CO$ is the most general operator of a fixed
dimension $m$.  To first order in conformal perturbation theory,
one has\rzamo
\eqn\eIIIii{
\d_\zb  J  = \lambda \oint_z  \frac{dw}{2\pi i }
{}~ \cos (\bh \Phi (w, \zb ) ) ~ J(z) . }
Having fixed $m$, based on the hypothesis of the last section one
expects that $J$ is conserved only for $\bh^2 = 2/(N+1)$, where
$m = N(n-1)$.  Since $n$ is odd
one has the following possibilities for the lowest dimensions of $\CO$:
\eqn\eIIIiii{
\eqalign{
{\rm dim} (\CO) &= 2: ~~~~~~~(N=1, n=3)  \cr
{\rm dim} (\CO) &= 4: ~~~~~~~(N=1, n=5), ~(N=2, n=3)  \cr
{\rm dim} (\CO) &= 6: ~~~~~~~(N=1, n=7), ~ (N=3, n=3).   \cr
{\rm dim} (\CO) &= 8: ~~~~~~~(N=1, n=9), ~(N=2, n=5) , ~
(N=4, n=3) .   \cr
{\rm dim} (\CO) &= 10: ~~~~~~(N=1, n=11), ~ (N=5, n=3).   \cr
{\rm dim} (\CO) &= 12: ~~~~~~(N=1, n=13), ~ (N=2, n=7) ,~
(N=3, n=5),~ (N=6, n=3) .   \cr
}}
The cases above at $N=1$ correspond to the bosonic form of the
lowest six higher charges at the free fermion point.  The additional
solutions at $N=2, 3, 4, 5,6$
correspond to the first higher charge at grade 3,
and the next higher charges at grades 5,7.

Consider first the simplest example where $\CO$ is of dimension 2.
The most general such operator is
\eqn\one{
\CO (z) =  a \d_z^2 \phi + b ( \d_z \phi )^2  . }
Since the two terms above can be related by a total derivative
of the form
$ \d_z ( \d_z \phi \exp (2i\phi /\bh ) )$, it is sufficient to
consider just one of them.  Take  $\CO = \d_z^2 \phi $.
According to \eIIIii, $J$ is conserved if the $1/(z-w)$ term
in the operator product expansion (OPE)  $\cos (\bh \Phi (w, \zb ) )
J(z)$ is a total derivative.   For $\CO$ of dimension 2,
only the $\exp (-i \bh \Phi )$ piece of $\cos \bh \Phi $
can give a pole, however as we will see for higher dimension
$\CO$ one must consider both pieces.   The relevant OPE is
\eqn\two{
e^{-i\bh \phi (z) } J(w)
\sim  \inv{z-w}
\( - \frac{\bh^2}{6} \phi''' + i (  \bh^3 /2 - \bh ) \phi' \phi''
+ \frac{\bh^4}{6}  ( \phi ' )^3 \)
e^{i (2/\bh - \bh ) \phi (w) } + ... }
where $\phi ' = \d_w \phi$ etc.
Above we have only written down the simple pole term since it is the only term
that is relevant for our analysis.
Integrating by parts, one finds that
the operator coefficient of the simple pole is
\eqn\ethree{
\d_w \[
\( - \frac{\bh^2}{6}  \phi'' + \frac{i}{6}  ( \bh^3 - 2\bh )
(\phi' )^2  \)
e^{i (2/\bh - \bh ) \phi (w)} \]
 + \frac{2}{3} ( \bh^2 -1 ) (\phi')^3
e^{i (2/\bh - \bh ) \phi (w)}  .   }
Thus, we only have a conserved current with $\CO$ of dimension 2 if
$\bh^2 = 1$, which is the case $(N=1, n=3)$ listed in \eIIIiii.

Consider next the case when $\CO$ is dimension 4.  After taking into
account total derivatives, the most general $\CO$ is
\eqn\ethreeb{
\CO = \d_z^4 \phi +  A (\d_z \phi )^4 . }
One needs the following OPE's:
\eqn\efour{\eqalign{
e^{-i\bh \phi (z)} \d_w^4 \phi (w) e^{2i\phi (w)/\bh }
&\sim \inv{z-w}
[
- \frac{\bh^2}{20}  \phi'''''  + i (\frac{\bh^3}{4} - \bh ) \phi' \phi''''
+ \frac{i}{2} \bh^3 \phi'' \phi'''
+ \frac{\bh^4}{2} (\phi')^2 \phi''' \cr
&~~~~~~
+ \frac{ 3\bh^4}{4} \phi' (\phi'')^2
- \frac{i}{2} \bh^5 (\phi')^3 \phi''  - \frac{\bh^6}{20} (\phi')^5
]
e^{i (2/\bh - \bh ) \phi (w)}   +.... \cr }}

\eqn\efive{\eqalign{
e^{-i\bh \phi (z)} (\d_w \phi )^4  e^{2i\phi (w)/\bh }
& \sim
\inv{z-w}
[ - \frac{i}{120} \bh^5 \phi'''''
+ \( \frac{\bh^4}{6} - \frac{\bh^6}{24} \) \phi' \phi''''
- \frac{\bh^6}{12} \phi''' \phi''
\cr
&~~~~~~~
+ i \( \bh^3 - \frac{2}{3} \bh^5 + \frac{\bh^7}{12} \)
(\phi')^2 \phi'''
+ i \( -\frac{\bh^5}{2} + \frac{\bh^7}{8} \) \phi' (\phi'')^2
\cr
&~~~~~~~
+ \( -2 \bh^2 + 3 \bh^4 - \bh^6 + \frac{\bh^8}{12} \)
(\phi')^3 \phi''
+ \cr
&~~~~~~~
i\( -\bh + 2\bh^3 - \bh^5 + \inv{6} \bh^7 - \inv{120} \bh^9 \)
(\phi')^5
]
e^{i (2/\bh - \bh ) \phi (w)}   +.... \cr }}
One finds that if
\eqn\esix{
A = - i \frac{ \bh^3 (3\bh^2 -10 ) }{2 (18 - 11 \bh^2 ) }  , }
then the current is conserved, so long as
\eqn\eseven{
\bh^2 (\bh^2 -1 )(\bh^2 +1 ) (\bh^2 -3 ) (3 \bh^2 -2 ) = 0.}
This yields three solutions for positive $\bh^2 = 2/3, 1, 3$.
Since conformal perturbation theory is only meaningful
when the perturbation is a relevant operator, which corresponds
to $\bh^2 < 2$,  we are left with the only two solutions listed
in \eIIIiii.

With increasing dimension of $\CO$, the above computations
rapidly become very complicated.  With the aid of Mathematica,
we have confirmed the predictions in \eIIIiii\ for
${\rm dim} \CO = 6$.  Repeating the above procedure, one finds
a conserved current provided $\bh^2$ is a solution of the equation:
\eqn\eight{
 -840 + 4978 \bh^2 - 11397 \bh^4  + 12862 \bh^6 - 7590 \bh^8
+ 2302 \bh^{10}  - 333 \bh^{12} + 18 \bh^{14}  = 0.
}
This polynomial can be factorized:
\eqn\enine{
(\bh^2 - 7) (\bh^2 -5 ) (\bh^2 -3) (\bh^2 - 1 )
(2\bh^2 - 1) (3\bh^2 - 4) (3\bh^2 - 2 ) = 0. }
After considering the OPE of the current with
the other piece $\exp ( i\bh \Phi )$  of the $\cos \bh \Phi$
perturbation, one finds that only $\bh^2 = 1/2, 1$  (in
the allowed range of $\bh^2$ ) remain as solutions, which again
are the two solutions listed in \eIIIiii.

\newsec{Concluding Remarks}

We have provided evidence for an $\hat{sl(2)}$ affine Lie
algebra symmetry in the sine-Gordon theory at the reflectionless
points.  The main difficulty which remains toward proving
this result is that  many of the relations in \eIIxii\ arise in
higher order conformal perturbation theory.

Assuming this affine symmetry exists, the interesting question
is to understand to what extent the symmetry characterizes
the main dynamical properties, such as the form factors and
correlation functions.  For example, it would be very interesting
to obtain the form factors  from  ordinary affine Lie algebraic
vertex operators.

\centerline{\bf Acknowledgements}

A.L. would like to thank I. Vaysburd for discussions.
A.L. is supported by an Alfred P. Sloan Foundation fellowship,
and the National Science Foundation in part through the
National Young Investigator program.

\listrefs

\end